\documentclass[final,5p,times,twocolumn]{elsarticle}

\usepackage{lineno}
\usepackage{graphicx}
\usepackage{amssymb}
\usepackage{pdflscape}
\usepackage[english]{babel}
\usepackage[dvipsnames,svgnames,x11names]{xcolor}
\usepackage{booktabs,tabularx}
\usepackage{multirow}
\usepackage{subfig} 
\usepackage{hyperref}
\usepackage{amsmath}
\usepackage{commath}
\usepackage[english]{babel}
\usepackage[ruled]{algorithm2e}
\SetEndCharOfAlgoLine{}
\usepackage[leftmargin=6em,rightmargin=12em,indentfirst=false]{quoting}
\usepackage{siunitx}
\usepackage{comment} 

\usepackage[final]{changes}
\usepackage{float}

\usepackage{lineno}
\usepackage{tikz}
\usepackage{comment} 
\usepackage[export]{adjustbox}

\usepackage{graphicx}
\usepackage[utf8]{inputenc}
\usepackage[export]{adjustbox}
\usepackage{wrapfig}
\usepackage{dcolumn}

\makeatletter
\newcolumntype{T}[3]{>{\textfont0=\the@{#1}{#2}{#3}}c<{\DC@end}}
\makeatother

\usepackage{pgfplots}
\pgfplotsset{width=10cm,compat=1.9}

\usepackage{array}

\newcolumntype{L}[1]{>{\raggedright\let\newline\\\arraybackslash\hspace{0pt}}m{#1}}
\newcolumntype{C}[1]{>{\centering\let\newline\\\arraybackslash\hspace{0pt}}m{#1}}
\newcolumntype{R}[1]{>{\raggedleft\let\newline\\\arraybackslash\hspace{0pt}}m{#1}}

\usepackage{subfiles}

\usepackage{todonotes}

\journal{Scientific Data}
\begin{document}
	
	\begin{frontmatter}
		\title{The Building Data Genome Project 2, energy meter data from the ASHRAE Great Energy Predictor III competition}
		
        \author{Clayton Miller$^{a,*}$, Anjukan Kathirgamanathan$^b$, Bianca Picchetti$^c$, Pandarasamy Arjunan$^d$, June Young Park$^e$, Zoltan Nagy$^e$, Paul Raftery$^f$, Brodie W. Hobson$^g$, Zixiao Shi$^g$, Forrest Meggers$^i$}

		\address{$^a$Building and Urban Data Science (BUDS) Lab, Dept. of Building, School of Design and Environment (SDE), National University of Singapore (NUS), Singapore}
		\address{$^b$UCD Energy Institute, University College Dublin, Ireland}
		\address{$^c$Gerencia Ciclode del Combustible Nuclear, Comisi\'{o}n Nacional de Energ\'{i}a At\'{o}mica, Buenos Aires, Argentina}
		\address{$^d$Berkeley Education Alliance for Research in Singapore (BEARS), Singapore}
		\address{$^e$Intelligent Environments Lab (IEL), Department of Civil, Architectural and Environmental Engineering, Cockrell School of Engineering, The University of Texas at Austin, TX, USA}
		\address{$^f$Center for the Built Environment, University of California, Berkeley, CA, USA}
		\address{$^g$Department of Civil and Environmental Engineering, Carleton University, ON, Canada}
		\address{$^i$CHAOS Laboratory, School of Architecture, Princeton University, Princeton, NJ, USA}
        \address{$^*$Corresponding Author: clayton@nus.edu.sg, +65 81602452}

%
\begin{abstract}
This paper describes an open data set of 3,053 energy meters from 1,636 non-residential buildings with a range of two full years (2016 and 2017) at an hourly frequency (17,544 measurements per meter resulting in approximately 53.6 million measurements). These meters were collected from 19 sites across North America and Europe, with one or more meters per building measuring whole building electrical, heating and cooling water, steam, and solar energy as well as water and irrigation meters. Part of these data were used in the Great Energy Predictor III (GEPIII) competition hosted by the ASHRAE organization in October-December 2019. GEPIII was a machine learning competition for long-term prediction with an application to measurement and verification. This paper describes the process of data collection, cleaning, and convergence of time-series meter data, the meta-data about the buildings, and complementary weather data. This data set can be used for further prediction benchmarking and prototyping as well as anomaly detection, energy analysis, and building type classification.

\end{abstract}

\end{frontmatter}


\section*{Background \& Summary}



Building performance analytics and commissioning processes have significant opportunities to save energy, reduce carbon emissions of buildings, and reduce the operating costs of building owners world-wide \cite{Kramer2019-lt}. Machine learning and prediction techniques are a vital component of many of the ways of finding savings opportunities and quantifying the risk and reward of undertaking such efforts. Despite the significant research body of knowledge developed, there is still a lack of understanding of how to scale techniques across the highly heterogeneous building stock \cite{Miller2019-sg}. When it comes to machine learning innovation in academia, one of the most significant assets can be large and open data sets that the community can use to prototype and quantitatively compare techniques in ways that show better value in terms of speed, accuracy, or implementation ease. This statement is supported by the significant efforts in time-series data classification \cite{Dau2018-ud}, image recognition \cite{Xiao2017-rd}, and the larger machine learning community in general, both hardware and software \cite{Mattson2020-ft}.\\

The building energy analytics community has only just started to use open data sets towards the efforts of creating benchmarking data sets. Several prominent open building energy-related data sets have been released in recent years including applications to building-level office \cite{Kriechbaumer2018-zg} and residential \cite{Kelly2015-uw} appliances, occupant behavior \cite{Mahdavi2019-wf}, heat pump \cite{Ruhnau2019-nx} and natural ventilation systems \cite{Schweiker2019-ro}, as well as commercial and residential energy meter data \cite{Rashid2019-pj, Paige2019-kx, Klemenjak2020-wv}. The use of open data sets in the built environment enables the analysis of large numbers of buildings in applications such as benchmarking \cite{Roth2020-ss}. From the machine learning perspective, there have also been efforts towards using large data sets to benchmark various machine learning techniques as applied to building energy performance analytics \cite{Granderson2016-wq}.\\

This paper focuses on the development of a data set that builds upon these motivations. The data set is part of the \emph{Building Data Genome Project}, an international consortium of building energy-related academics and practitioners who seek to create large, open data sets that increase the understanding of the foundations of building behavior and energy use in buildings. The first phase of the project had a data set that was released in 2017 and included one year of hourly data from over 500 buildings \cite{Miller2017-fq}. \\

\begin{figure*}[ht!]
\includegraphics[width=\textwidth,height=\textheight,keepaspectratio]{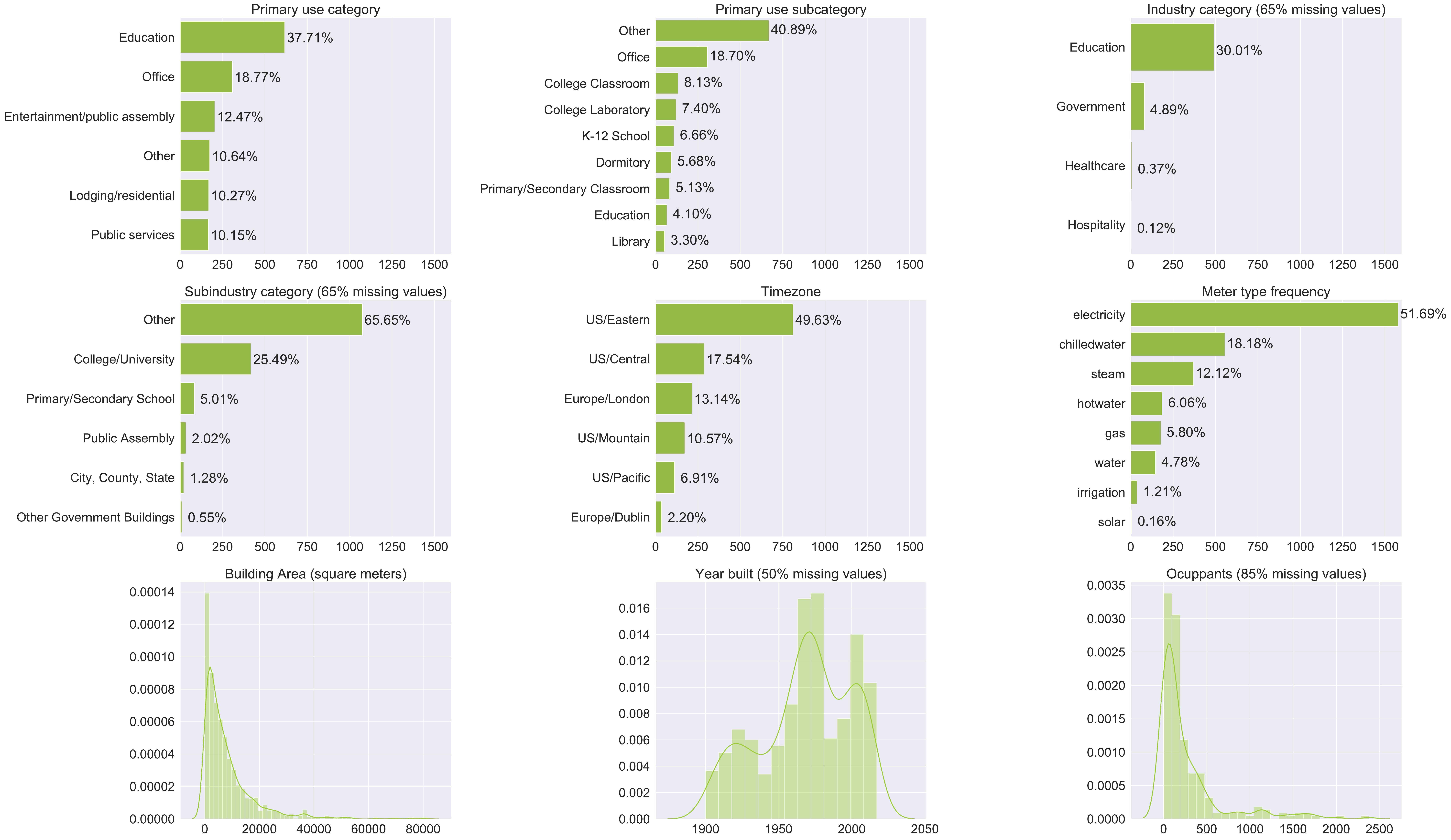}
\caption{Main features distribution in metadata file that describes the various buildings from which the meter data was collected. Several of the meta-data categories are available for all buildings including the Primary Use Category of the building (\texttt{primaryspaceusage}), the Sub-primary Use Category (\texttt{subprimaryspaceusage}), Gross Floor Area (\texttt{sqm}), Time Zone (\texttt{timezone}), Weather Data, and Meter Type.}
\label{tab:fig-metadata}
\end{figure*}

The newest version of the data set is described in this publication as the Building Data Genome Project 2 (BDG2) data set. This open data repository has data from 1,636 non-residential buildings. It includes hourly whole-building data for two years, from different kinds of meters: electricity, chilled water, steam, hot water, gas, water, irrigation, and solar. The hourly frequency for the data set was targeted as it provides enough resolution to support analytics techniques targeting several scales, including daily, weekly, monthly, seasonal, and annual patterns of use. Each of the buildings has metadata such as area, weather, and primary use type collated. This data set can be used to benchmark various statistical learning algorithms and other data science techniques. It can also be used merely as a teaching or learning tool to practice dealing with measured performance data from large numbers of non-residential buildings. This data set was collected from 19 different locations from around the world. These locations, climates, and the number of buildings from each site are found in Table \ref{tab:sites-tab1}. This table also includes information about which buildings were used in the ASHRAE-sponsored Great Energy Predictor III (GEPIII) competition that was held on the Kaggle platform from October to December 2019 (\url{https://www.kaggle.com/c/ashrae-energy-prediction}) \cite{gepIII-overview}. These buildings represent several different primary use type categories from several industries. Figure \ref{tab:fig-metadata} illustrates the breakdown of the buildings according to the principal use category and subcategory, industry and sub-industry, timezone, and meter type. The remaining parts of this paper focus on how the data were collected, processed, and how users can find and use the data for several example applications.

\begin{table*}[t]
\caption{Overview of the sites from which the building energy meter data was collected. Each site is given an animal-like site code name, a UID that corresponds to some of the data convergence processes, the Kaggle Site ID that was included in the competition, and the Actual Site Name, Location and Climate Zone. Several of the sites are to remain anonymous based on discussions with the data donors. The last two columns indicate the number of buildings and meters where two years of hourly, whole building meter data were collected from each site. The climate zones labels are from the ASHRAE climate classification system.}
\label{tab:sites-tab1}
\centering
\begin{adjustbox}{width=1\textwidth}
\begin{tabular}{|l|c|c|l|l|c|r|r|}
\hline
\textbf{Site} & \textbf{UID} & \textbf{Kaggle} & \textbf{Actual Site Name} & \textbf{Location} & \textbf{Climate} & \textbf{Buildings} & \textbf{Meters} \\
\hline
Panther & 1P4YFG & 0 & Univ. of Central Florida (UCF) & Orlando, FL & 2A & 136 & 299 \\
\hline
Robin & 1TKL5P & 1 & Univ. College London (UCL) & London, UK & 4A & 52 & 67 \\
\hline
Fox & 4QFLSM & 2 & Arizona State Univ. (ASU) & Tempe, AZ & 2B & 137 & 306 \\
\hline
Rat & 72SGIQ & 3 & Washington DC - City Buildings & Washington DC & 4A & 305 & 305 \\
\hline
Bear & 7E44IQ & 4 & Univ. of California - Berkeley & Berkeley, CA & 3C & 92 & 92 \\
\hline
Lamb & 9T5ZA2 & 5 & Cardiff - City Buildings & Cardiff, UK & 4A & 147 & 265 \\
\hline
Eagle & EQDHIP & 6 & Anonymous & N/A & 4A & 47 & 106 \\
\hline
Moose & H7PNXU & 7 & Ottawa - City Buildings & Ottawa, Ontario & 6A & 15 & 43 \\
\hline
Gator & I9U4WZ & 8 & Anonymous & N/A & 2A & 74 & 74 \\
\hline
Bull & JG98YH & 9 & Univ. of Texas - Austin & Austin, TX & 2A & 124 & 308 \\
\hline
Bobcat & JP4TNW & 10 & Anonymous & N/A & 5B & 36 & 116 \\
\hline
Crow & JTM0LY & 11 & Carleton Univ. & Ottawa, Ontario & 6A & 5 & 15 \\
\hline
Wolf & RFO3TV & 12 & Univ. College Dublin (UCD) & Dublin, Ireland & 5A & 36 & 66 \\
\hline
Hog & SREOJG & 13 & Anonymous & Anonymous & 6A & 163 & 336 \\
\hline
Peacock & WI83D6 & 14 & Princeton University & Princeton, NJ & 5A & 106 & 298 \\
\hline
Cockatoo & YYAFES & 15 & Cornell University & Cornell, NY & 6A & 124 & 282 \\
\hline
Shrew & L2HJLD & - & UK Parliment & London, UK & 4A & 9 & 13 \\
\hline
Swan & N950XM & - & Anonymous & N/A & 3C & 21 & 55 \\
\hline
Mouse & ZVJUMW & - & Ormand Street Hospital & London, UK & 4A & 7 & 7 \\
\hline
\end{tabular}

\end{adjustbox}
\end{table*}

\section*{Methods}

\subsection*{Energy data sources overview and collection}
The collection of the metadata and whole building meter data from the various sites outlined in Table \ref{tab:sites-tab1} was done by the authors of this paper from September 2017 until May of 2019. Eight of the sites from this list are online data sources that are freely downloadable without the use of login credentials. These sites are considered open access data sources and are publicly available. Table \ref{tab:sites-tab2} outlines these eight sites and the online link to the main interface for downloading the data. The remaining eleven sites did not have online, publicly available data feeds. In those situations, there were facilities management professionals involved in the process of data collection and organization for those subsets. Data collection from these sites was a manual process that included site visits, in-person meetings, and data collection workshops, numerous digital communications via video calls and emails. The raw meter data for these sites were downloaded and provided to the technical team, usually via emailing flat files. These raw data sources are not included in the data repository; however, the process of convergence, cleaning, and normalization is included in this paper's subsequent subsections.

\begin{table*}[t]
\caption{Sites with data that are publicly available to download online. The site name includes the Kaggle ID in parentheses.}
\label{tab:sites-tab2}
\centering
\begin{adjustbox}{width=1\textwidth}
\begin{tabular}{|l|l|l|}
\hline
\textbf{Site} & \textbf{Actual Site Name} & \textbf{Online source}  \\
\hline
Panther (0) & Univ. of Central Florida (UCF) & \url{http://oeis.ucf.edu/} \\
\hline
Robin (1) & Univ. College London (UCL) & \url{https://platform.carbonculture.net/communities/ucl/30/} \\
\hline
Fox (2) & Arizona State Univ. (ASU) & \url{https://cm.asu.edu/} \\
\hline
Bear (4) & UC Berkeley (UCB) & \url{https://engagementdashboard.com/ucb/ucb/} \\
\hline
Lamb (5) & Cardiff - City Buildings & \url{https://platform.carbonculture.net/communities/cardiff-council/19/} \\
\hline
Cockatoo (15) & Cornell University & \url{https://portal.emcs.cornell.edu/} \\
\hline
Shrew & UK Parliment & \url{https://platform.carbonculture.net/communities/uk-parliament/2/} \\
\hline
Mouse & Ormand Street Hospital & \url{https://platform.carbonculture.net/communities/great-ormond-street-hospital/4/} \\
\hline
\end{tabular}                

\end{adjustbox}
\end{table*}

\subsection*{Weather data overview and collection}
\begin{figure*}[ht!]
\includegraphics[width=\textwidth,height=\textheight,keepaspectratio]{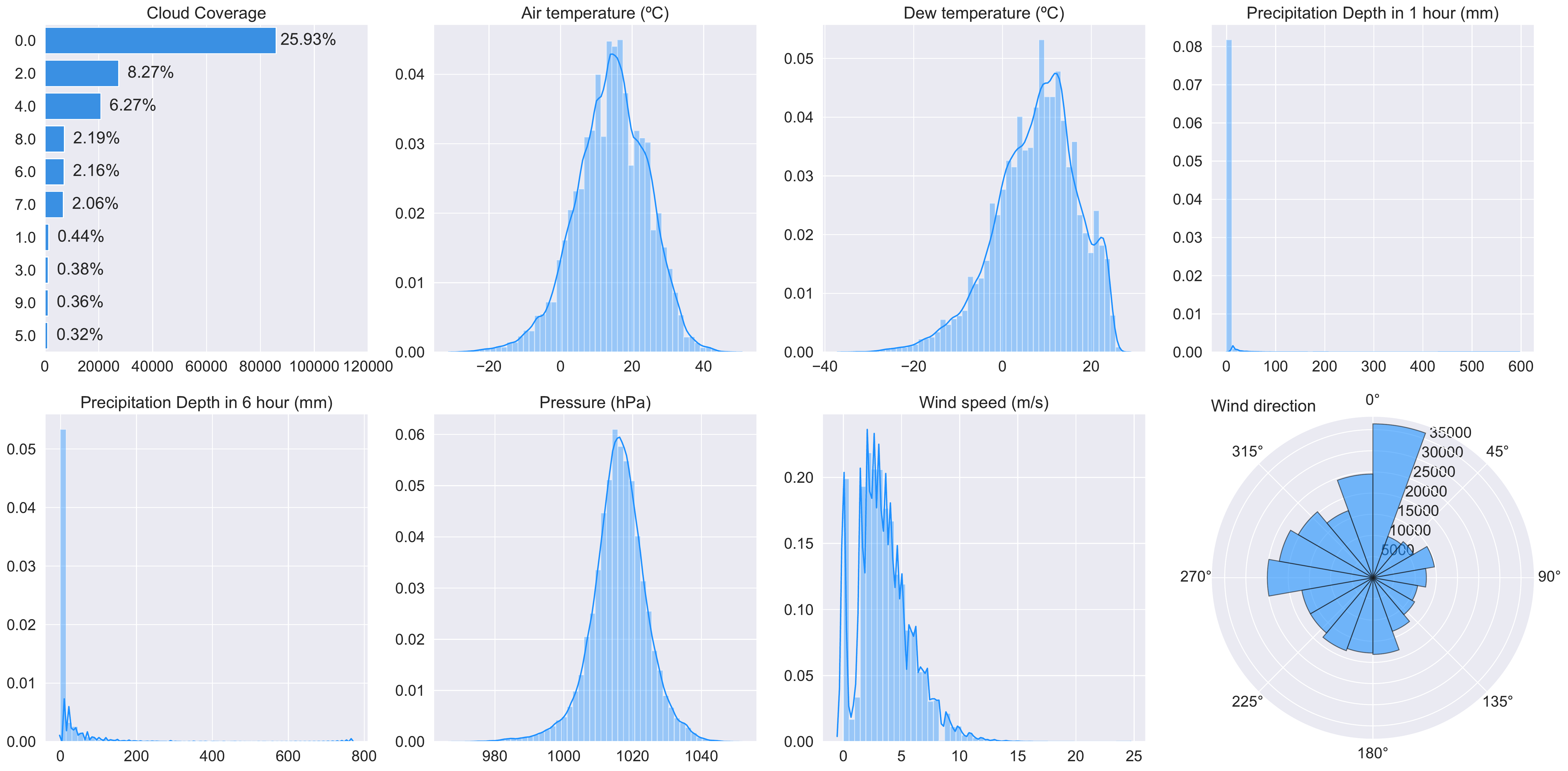}
\caption{Main feature distributions of the weather data set.}
\label{tab:fig-weather}
\end{figure*}

One of the critical comparative data sources for building energy meter data is outside weather conditions, which are among the key influencing factors for energy consumption in buildings. Each of the building sites has a corresponding weather data file with hourly data related to the outdoor temperature, humidity, cloud cover, and other conditions that influence energy consumption. Hourly weather data for this data set were collected using the National Centers for Environmental Information (NCEI) National Oceanic and Atmospheric Administration (NOAA) Integrated Surface Database (ISD) (\url{https://www.ncdc.noaa.gov/isd}). The ISD-Lite version was used for easy hourly data capture. The closest station with available data for the period 2016-2017 was selected for each site, as outlined in Table \ref{tab:sites-tab3}. The ISD-Lite data set includes the eight climatological variables for each station with a modified timestamp, which corresponds to the nearest hour of actual observation. In the preparation step for this data set, scaling (where applied) was removed and missing values were processed to be \texttt{NaN} instead of \texttt{-9999} as per the raw data. The final processed weather data is summarised in Figure \ref{tab:fig-weather}.

\begin{table}[t]
\caption{ISD weather station data sources for the non-anonymous sites. The site name includes the Kaggle ID in parentheses.}
\label{tab:sites-tab3}
\centering
\begin{tabular}{|l|c|c|l|l|c|r|}
\hline
\textbf{Site} & \textbf{ISD Station Code} \\
\hline
Panther (0) & 722050-12815 \\
\hline
Robin (1) & 037720-99999 \\
\hline
Fox (2) & 722780-23183 \\
\hline
Rat (3) & 724050-13743 \\
\hline
Bear (4) & 724930-23230 \\
\hline
Lamb (5) & 037150-99999 \\
\hline
Moose (7) & 710630-99999 \\
\hline
Bull (9) & 722544-13958 \\
\hline
Crow (11) & 710630-99999 \\
\hline
Wolf (12) & 039690-99999 \\
\hline
Peacock (14) & 724095-14792 \\
\hline
Cockatoo (15) & 725155-94761 \\
\hline
Shrew & 037720-99999 \\
\hline
Mouse & 037720-99999 \\
\hline
\end{tabular}                
\end{table}

\subsection*{Data cleaning and normalization}
After collection of the raw data from the sites and weather sources, the data were transformed in ways that create consistency and uniformity across the data sets so that they could be converged into one large data set. These steps were completed in a private, non-public data repository as the preparation for the data was done in the Kaggle GEPIII competition context. These data and processes were kept secure as the premature release of the data would have compromised the competition's integrity. This subsection describes those steps used to create both the data set for the competition and this data repository.\\

The first step in this process was the normalization of measurement units for the various energy meter types. Table \ref{tab:sites-tab4} summarises the original measurement units for the raw data collected from every site. A conversion process was undertaken to convert into standard units for every meter type, as outlined in Table \ref{tab:sites-tab5}. Following the standardization of the units, a few additional steps were undertaken to clean and process the data. All meters with only a single value were removed, duplicate meter data (if present) were removed, and negative meter values were replaced with \texttt{NaN}. Where there were more than 50\% of negative meter readings, this meter was also removed. This step removes the possibility of including meters from net-zero energy buildings, although we are not aware that there were any of these buildings in the data set specifically. Meters with significant consecutive missing values (over 100 consecutive days) were excluded. There were still meters with very high-value outliers, and in this case, standard outlier removal techniques won't work as these outliers are large enough to skew measures such as the mean. A log conversion and pruning technique were used with any high outliers greater than three standard deviations from the mean on the log-transformed data converted to \texttt{NaN} values. Finally, all meter data was rounded to four decimal places. \\

For the metadata of the buildings, where necessary, floor area in \texttt{sqm} and \texttt{sqft} was converted from whichever floor area data was available. Latitude and longitude data were set to the central location of either the site or the city in which the site is located. In all cases, all buildings are within a 25-mile (40-kilometer) radius of the central location of the site or city. For the \texttt{year$\_$built} attribute, a valid range was considered to be 1900 to 2018, and invalid or implausible years were filled as missing values. Primary space usage (\texttt{primary$\_$use}) metadata for all buildings was mapped using the Energy Star scheme building description types. Based upon the meter and metadata as described above, a further filter was done to synchronize both sets of data and remove meter data for which the building metadata did not exist and likewise remove metadata for which meter data did not exist.

\begin{table*}[t]
\caption{Overview of original measurement units for the raw data collected from each site. All data were subsequently converted to kWh$_{sum}$ or liters. The site name includes the Kaggle ID in parentheses.} 

\label{tab:sites-tab4}
\centering
\begin{adjustbox}{width=1\textwidth}
\begin{tabular}{|l|c|c|c|c|c|c|c|c|}
\hline
\textbf{Site} & \textbf{Chilled water} & \textbf{Electricity} & \textbf{Gas} & \textbf{Hot water} & \textbf{Solar} & \textbf{Steam} & \textbf{Water} & \textbf{Irrigation} \\
\hline
Panther (0) & kBTU$_{sum}$ & kBTU$_{sum}$ & kBTU$_{sum}$ & & & & gallons & gallons \\
\hline
Robin (1) & & kWh$_{sum}$ & & & & & & \\
\hline
Fox (2) & Tons$_{avg}$ & kW$_{avg}$ & & mmBTU$_{sum}$ & & & & \\
\hline
Rat (3) & & kWh$_{sum}$ & & & & & & \\
\hline
Bear (4) & & kW$_{avg}$ & & & & & & \\
\hline
Lamb (5) & & kWh$_{sum}$ & kWh$_{sum}$ & & & & & \\
\hline
Eagle (6) & mmBTU$_{sum}$ & kW$_{avg}$ & & mmBTU$_{sum}$ & & lbs$_{perhour}$ & & \\
\hline
Moose (7) & MJ & MJ & & & & & & \\
\hline
Gator (8) & & kWh$_{avg}$ & & & & & & \\
\hline
Bull (9) & Tons$_{sum}$ & kWh$_{sum}$ & & & & lbs$_{perhour}$ & & \\
\hline
Bobcat (10) & kBTU & kWh$_{sum}$ & kBTU & kBTU & kWh$_{sum}$ & & gallons & \\
\hline
Crow (11) & kWh$_{sum}$ & kWh$_{sum}$ & kWh$_{sum}$ & & & & & \\
\hline
Wolf (12) & & kWh$_{sum}$ & m3 & & & & liters & \\
\hline
Hog (13) & Tons$_{avg}$ & kWh$_{avg}$ & & & & lbs$_{perhour}$ & & \\
\hline
Peacock (14) & Tons$_{avg}$ & kW$_{avg}$ & & & & lbs$_{perhour}$ & & \\
\hline
Cockatoo (15) & Tons$_{avg}$ & kW$_{avg}$ & & Tons$_{avg}$ & & lbs$_{perhour}$ & & \\
\hline
Shrew & & kWh$_{sum}$ & kWh$_{sum}$ & & & & & \\
\hline
Swan & Tons$_{avg}$ & kWh$_{sum}$ & & kBTU$_{sum}$ & & & & \\
\hline
Mouse & & kWh$_{sum}$ & & & & & & \\
\hline
\end{tabular}           

\end{adjustbox}
\end{table*}

\begin{table}[t]
\caption{Overview of measurement unit conversion process. All energy-related meters were converted to \emph{kWh$_{sum}$ or \emph{liters}} from the various raw data units}
\label{tab:sites-tab5}
\centering
\begin{tabular}{|c|c|}
\hline
\textbf{Unit} & \textbf{Conversion Factor} \\
\hline
kW$_{avg}$ & 1 kWh$_{sum}$ = kW$_{avg}$ * 1 \\
\hline
tons & 1 kWh$_{sum}$ = tons * 3.51685 \\
\hline
kBTU & 1 kWh$_{sum}$ = kBTU * 0.293071 \\
\hline
MJ & 1 kWh$_{sum}$ = MJ * 0.277778 \\
\hline
mmBTU & 1 kWh$_{sum}$ = mmBTU * 293.071 \\
\hline
therm & 1 kWh$_{sum}$ = therm * 29.3071 \\
\hline
cubic meter gas & 1 kWh$_{sum}$ = cubic meter * 11.4772 \\
\hline
lb/hour steam & 1 kWh$_{sum}$ = lb/hour * 0.305 \\
\hline
gallons & 1 liter = gallons * 0.264172 \\
\hline
\end{tabular}           

\end{table}

\section*{Data Records}
This section documents the file types and structure for the data set (\url{https://github.com/buds-lab/building-data-genome-project-2}). The following subsections outline the data files that can be found in the repository to guide their use. Each building in the data set can be connected to this publication through its \emph{Unique Site Identifier} that was created with the following structure: \emph{animal name} (unique per site) + \emph{primary space usage abbreviation} + \emph{Human-like name} (unique per building). An example of a building name is \emph{Raven\_Education\_Nina}.\\

\subsection*{Building Metadata}
The building meta data file (\texttt{data/metadata/metadata.csv}) contains information about the whole building characteristics that enable the analysis of the associated meter data with various aspects of the building such as floor area, weather, and primary use type. These data were collected either from the operations teams from which the data was collected or from descriptors from the online data portals if collected from a public data source. Only the attributes for building unique identifier (\texttt{building\_id}), site identifier (\texttt{site\_id}), floor area (\texttt{sqft} and \texttt{sqm}), and time zone (\texttt{timezone}) are found for all the buildings. The remaining meta data descriptors have missing value rates from 4-99\%. A more detailed overview of these attributes can be found in the repository documentation (\url{https://github.com/buds-lab/building-data-genome-project-2/wiki/Metadata-description}).  The following are the attributes or column headings and the description of the data found in the file:

\begin{itemize}
\item
  \texttt{building\_id}: building code-name with the structure -
  \emph{ \_UniqueSiteID \_primaryspaceusage \_UniqueFirstName}.
\item
  \texttt{site\_id}: animal-code-name for the site.
\item
  \texttt{primaryspaceusage}: Primary space usage of all buildings is mapped using the Energy Star scheme building description types as seen Table \ref{tab:sites-tab6}.
\item
  \texttt{sqft}: Floor area of building in square feet (sq ft).
\item
  \texttt{lat}: Latitude of building location to city level. This attribute is available for all non-anonymous locations.
\item
  \texttt{lng}: Longitude of building location to city level.This attribute is available for all non-anonymous locations.
\item
  \texttt{electricity}: Presence of this kind of meter in the building. \texttt{Yes} if affirmative, \texttt{NaN} if negative.
\item
  \texttt{hotwater}: Presence of this kind of meter in the building. \texttt{Yes} if affirmative, \texttt{NaN} if negative.
\item
  \texttt{chilledwater}: Presence of this kind of meter in the building. \texttt{Yes} if affirmative, \texttt{NaN} if negative.
\item
  \texttt{steam}: Presence of this kind of meter in the building. \texttt{Yes} if affirmative, \texttt{NaN} if negative.
\item
  \texttt{water}: Presence of this kind of meter in the building. \texttt{Yes} if affirmative, \texttt{NaN} if negative.
\item
  \texttt{irrigation}: Presence of this kind of meter in the building. \texttt{Yes} if affirmative, \texttt{NaN} if negative.
\item
  \texttt{solar}: Presence of this kind of meter in the building. \texttt{Yes} if affirmative, \texttt{NaN} if negative.
\item
  \texttt{gas}: Presence of this kind of meter in the building. \texttt{Yes} if affirmative, \texttt{NaN} if negative.
\item
  \texttt{yearbuilt}: Year corresponding to when building was first constructed, in the format \texttt{YYYY}.
\item
  \texttt{numberoffloors}: Number of floors corresponding to building.
\item
  \texttt{date\_opened}: Date building was opened for use, in the format \texttt{D/M/YYYY}.
\item
  \texttt{sub\_primaryspaceusage}: Energy Star scheme building description types subcategory.
\item
 \texttt{energystarscore}: Rating of building corresponding to building Energy Star scheme\footnote{\url{https://www.energystar.gov/buildings/facility-owners-and-managers/existing-buildings/use-portfolio-manager/understand-metrics/how-1-100}}.
\item
  \texttt{eui}: Energy use intensity of the building (kWh/year/m2)\footnote{\url{https://www.energystar.gov/buildings/facility-owners-and-managers/existing-buildings/use-portfolio-manager/understand-metrics/what-energy}}.
\item
  \texttt{heatingtype}: Type of heating in corresponding building.
\item
  \texttt{industry}: Industry type corresponding to building.
\item
  \texttt{leed\_level}: LEED rating of the building\footnote{\url{https://www.usgbc.org/leed/}}.
\item
  \texttt{occupants}: Design condition number of occupants in the building.
\item
  \texttt{rating}: Other building energy ratings.
\item
  \texttt{site\_eui}: Energy (Consumed/Purchased) use intensity of the site
  (kWh/year/m2).
\item
  \texttt{source\_eui}: Total primary energy consumption normalized by area (Takes into account conversion efficiency of primary energy into secondary energy).
\item
  \texttt{sqm}: Floor area of the building in squared meters.
\item
  \texttt{subindustry}: More detailed breakdown of Industry type corresponding to building.
\item
  \texttt{timezone}: Site time zone.
\end{itemize}

\subsection*{Weather Data}
The building weather data file (\texttt{data/weather/weather.csv}) contains the time-series data for each building as it corresponds to the energy meters. These data have a time range from January 1, 2016, to December 31, 2017 - the same as the meter data files. A more detailed overview of these data can be found in the repository documentation (\url{https://github.com/buds-lab/building-data-genome-project-2/wiki/Weather-Description}). The following are the attributes or column headings and the description of the data found in the file: 

\begin{itemize}
\item
  \texttt{timestamp}: Date and Time in the format \texttt{YYYY-MM-DD hh:mm:ss} in the local timezone.
\item
  \texttt{site\_id}: human name-animal-code-name unique identifier for the site.
\item
  \texttt{airTemperature}: The temperature of the air in degrees Celsius (ºC).
\item
  \texttt{cloudCoverage}: Portion of the sky covered in clouds, in oktas (\url{https://en.wikipedia.org/wiki/Okta}).
\item
  \texttt{dewTemperature}: The dew point (the temperature to which a given parcel of air must be cooled at constant pressure and water vapor content for saturation to occur) in degrees Celsius (ºC).
\item
  \texttt{precipDepth1HR}: The depth of liquid precipitation measured over a one hour accumulation period (mm).
\item
  \texttt{precipDepth6HR}: The depth of liquid precipitation that is measured over a six-hour accumulation period (mm).
\item
  \texttt{seaLvlPressure}: The air pressure relative to Mean Sea Level (MSL) (mbar or hPa).
\item
  \texttt{windDirection}: The angle, measured in a clockwise direction, between true north and the direction from which the wind is blowing (degrees).
\item
  \texttt{windSpeed}: The rate of horizontal travel of air past a fixed point in (m/s).
\end{itemize}

\subsection*{Meter Data}
There are three sets of meter data found in the repository. The first is the \emph{raw} data set that includes the most substantial data set that was formed after convergence of the data from each source and the initial cleaning, unit conversion, and other processing steps outlined in the Data Cleaning and Normalization Section. The \emph{cleaned} data set provides a data set with another phase of cleaning and processing described below. Finally, there is a data set that includes the 2017 data that matches with the \emph{Kaggle} competition. This data set is included as several updates and conversions were performed on the BDG data sets after the competition. An overview of the differences between these data sets can be found in the repository documentation (\url{https://github.com/buds-lab/building-data-genome-project-2/wiki/Meters-data-features}).

\subsubsection*{Raw Meter Data}
There are eight files containing the time-series data for each building meter type. These files contain a column for each building in the data set for that particular meter. These files are contained in the \texttt{/data/meters/raw/} folder and includes the files \texttt{electricity.csv}, \texttt{hotwater.csv}, \texttt{chilledwater.csv}, \texttt{steam.csv}, \texttt{water.csv}, \texttt{irrigation.csv}, \texttt{solar.csv} and \texttt{gas.csv}. Each data file contains the data timestamp as the initial row in the format \texttt{YYYY-MM-DD hh:mm:ss} in the local timezone and one column per building in the data set in the units \emph{kWh$_{sum}$} for the energy-related meters and \emph{liters} for the non-energy meters. Each row represents one hour, and the reading is the energy or water sum across that hour. These data have a time range from January 1, 2016, to December 31, 2017 - the same as the weather data files. A more detailed overview of these data can be found in the repository documentation.

\subsubsection*{Cleaned Meter Data}
This folder content and structure (\texttt{/data/meters/cleaned/}) is similar to the \emph{raw} data folder, however, more outliers have been removed using the Twitter \emph{AnomalyDetection} R library (\url{https://github.com/twitter/AnomalyDetection}), zero readings longer than 24 continuous hours are removed, and zero readings in electricity meters are removed.

\subsubsection*{Kaggle Public test/validation Data}
This folder (\texttt{/data/meters/kaggle/}) includes a single file that contains the 2017 data of all the meters and sites from the GEPIII competition that was used as the public test/validation data set. This file can be used by those seeking to make a comparison to the training data found provided by the competition website. It can be used to train models and make submissions for the final score test data set (private leaderboard). This data set is provided as the other BDG2 data sets have been transformed since the competition. This original form allows users not to have to reverse those transforms to use the data in the competition. More details of the connection from this repository and the competition can be found in the Usage Notes section.

\section*{Technical Validation}
To illustrate to potential users the usefulness of the BDG2 data set, several data quality screening techniques have been applied to the time-series meter data to show an overview of the normalized consumption patterns across the data set, the completeness and quality of the data, the relationship between the weather and meter data, and the volatility of the data in terms of shifts in steady-state. Each of these screening techniques was developed and applied to the previous BDG1 data set in earlier work \cite{Miller2017-hp}. These screening techniques are designed to validate the technical capacity for the data sets to meet the needs of various applications. A more detailed overview of the screening process can be found in the repository documentation and in the Usage Notes section (\url{https://github.com/buds-lab/building-data-genome-project-2/wiki/Meters-data-screening}).

\subsection*{Normalized Consumption}
\begin{figure*}[ht!]
\centering
\includegraphics[width=0.85\textwidth,height=\textheight,keepaspectratio]{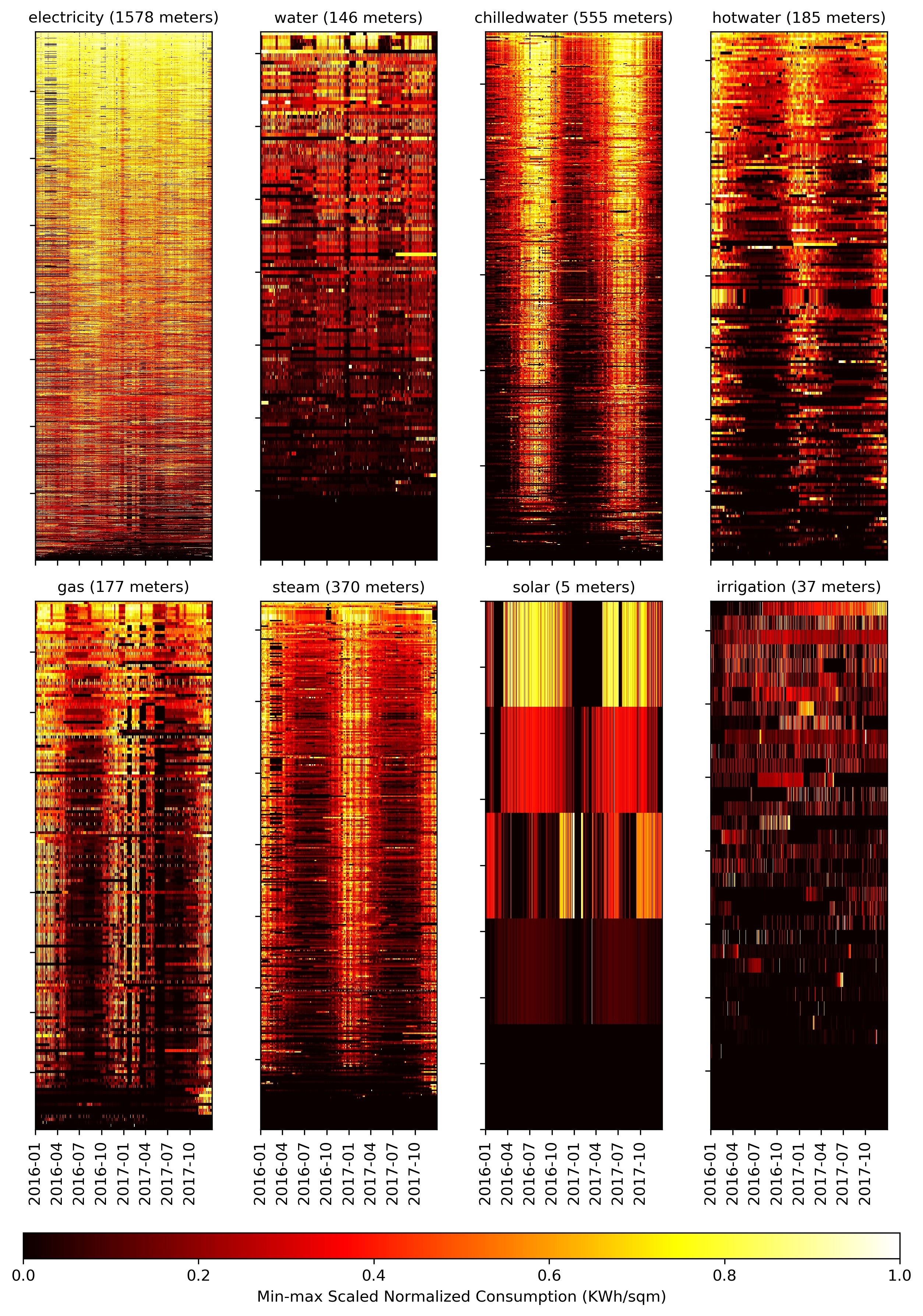}
\caption{Normalized meter consumption expressed as the daily energy consumption (kWh) or water consumption (liters) per area unit (square meters) of the building that is then scaled to Min-max scaling (for a range of 0-1). Each heatmap corresponds to a meter type, the horizontal access for all graphics is the two year time range, and the vertical axis are the range of meters sorted anonymously from (bottom-to-top) from lowest to highest scaled daily normalized consumption.}
\label{tab:fig-consumption}

\end{figure*}

The first screening technique applied is to visualize the meter data from a high level in a normalized way to see the general patterns and fluctuations across the data. The first step in this process is the summation of the hourly data across each day. The daily totals are then normalized once by dividing by the floor area (\texttt{sqm}) and then normalized again by scaling to the maximum and minimum for the time range for each meter data set. Figure \ref{tab:fig-consumption} illustrates the panel of the eight-meter types with this screening process applied. This figure illustrates each meter type in its own heat map where the horizontal axis for each heatmap is the two year period, and the vertical axis represents all of the meters for each category sorted from top to bottom according to the metric. This visualization technique is used in Figures \ref{tab:fig-consumption}-\ref{tab:fig-breakout}. For the normalized energy consumption technical validation, the various meters have seasonal, cyclical patterns that are apparent for a certain range of each meter type.

\subsection*{Data Quality}
\begin{figure*}[ht!]
\centering
\includegraphics[width=0.85\textwidth,height=\textheight,keepaspectratio]{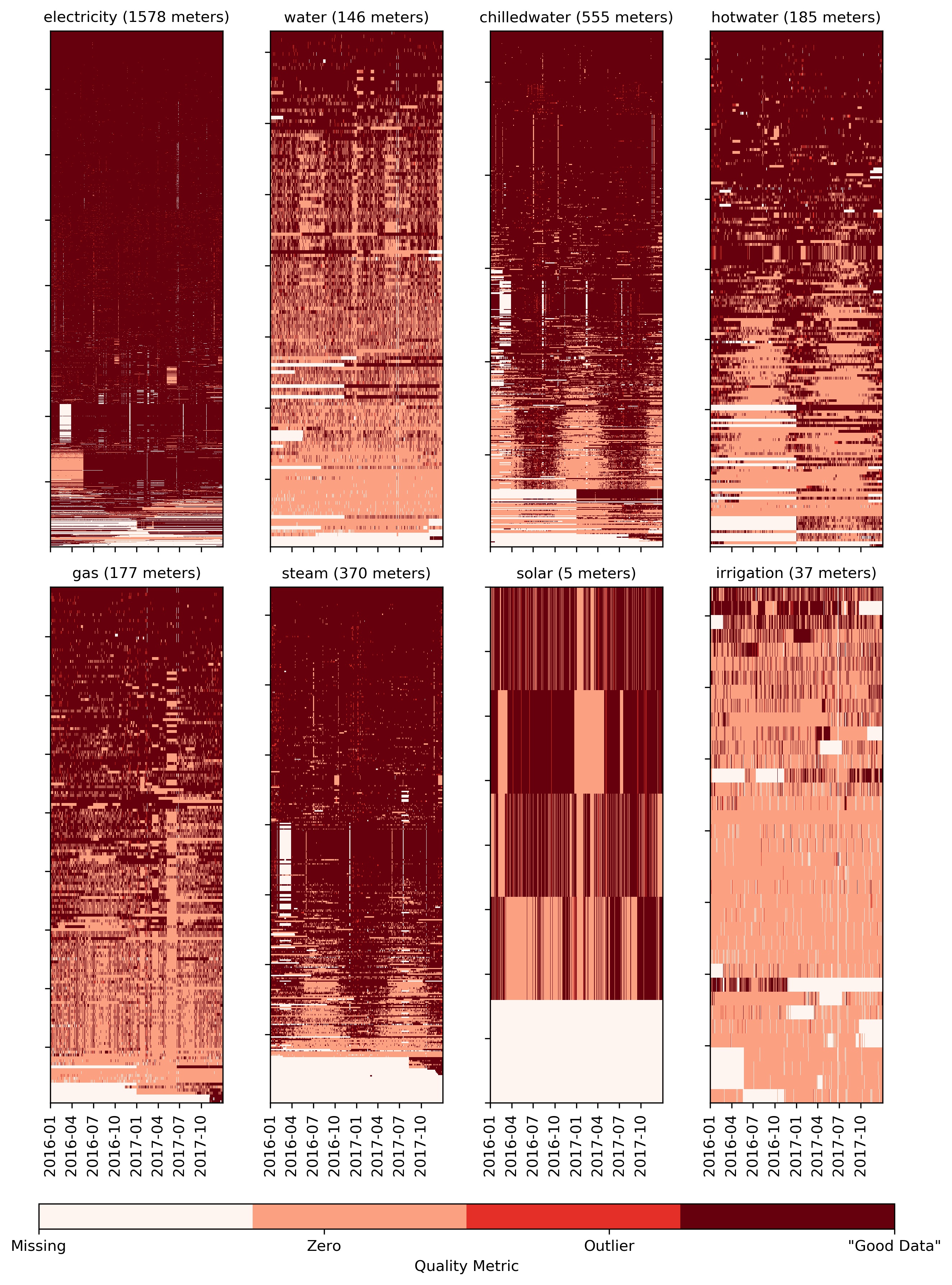}
\caption{Data quality plot of each meter type. Sorted (bottom-to-top) according to increasing number of \emph{good data}.}
\label{tab:fig-dataQuality}
\end{figure*}

The next screening technique applied is a set of filters applied to the time-series data from the meters to categorize four different types of readings of the data: missing data, data with a reading of zero, outliers, and the remaining data that can be considered the most informational (labeled as \emph{Good Data}). This process was applied to all the meter data sets, as shown in Figure \ref{tab:fig-dataQuality}. The outliers for the heat map are calculated using the Twitter \emph{AnomalyDetection} R library (\url{https://github.com/twitter/AnomalyDetection}). The resultant heat maps show a small percentage of the meters have a significant amount of missing data in certain time frames. These gaps are considered normal in meter data sets and can be the result of numerous technical or data collection issues. These data may also mean that the building was offline during certain periods. The meters still met the criteria to be included in the data set despite these gaps; therefore, the gaps are under a certain percentage of the overall data set as defined in earlier sections. The visualization also shows that there are a significant amount of zero readings for certain meter types, such as those related to heating, cooling, and irrigation. These zero measurement values make sense in those contexts, and these data are likely to be useful as they demonstrate periods when those systems are not in use. The screening shows few outliers as most of those data were filtered in previous cleaning steps outlined previously.

\subsection*{Weather Data Sensitivity}
\begin{figure*}[ht!]
\centering
\includegraphics[width=0.85\textwidth,height=\textheight,keepaspectratio]{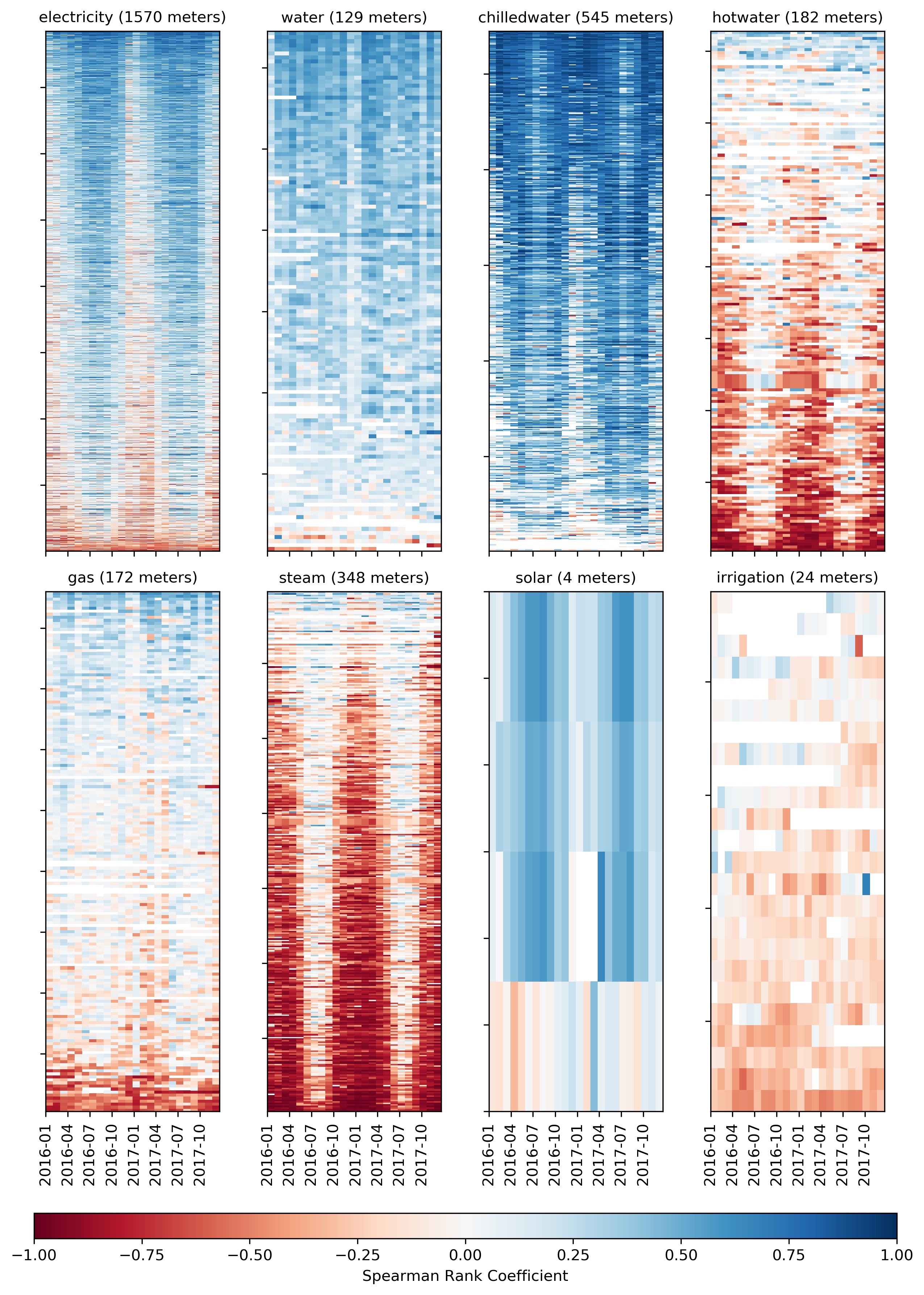}
\caption{Weather sensitivity plot of each meter type. Spearman rank coefficient was calculated between the meter reading (kWh or Liters) and the outside air temperature (Degrees Celsius) for each month. Sorted (bottom-to-top) according to increasing sum of coefficients.}
\label{tab:fig-weathersens}

\end{figure*}

The next screening process illustrates and validates the relationship between the meter data and the associated weather data files that are included in the repository. This validation step shows the value of providing these data sets in tandem and the influence of weather on buildings' energy consumption. This metric is calculated by taking a cleaned version of the data set in which days with only zero readings are removed and finding the Spearman rank-order correlation coefficient between the meter reading and the outside air temperature across each month. The resultant heat map visualization can be found in Figure \ref{tab:fig-weathersens}. The Spearman coefficient is a standard non-parametric measure of rank correlation. It shows which meter types are heavily positively correlated (related to cooling system energy influence) or negatively correlated (heating system energy influence). These heat maps illustrate the range of behavior for the various meters; the hot and chilled water and steam meters are heavily correlated, as expected, but also a significant number of electricity meters.

\subsection*{Breakout Detection}
\begin{figure*}[ht!]
\centering
\includegraphics[width=0.85\textwidth,height=\textheight,keepaspectratio]{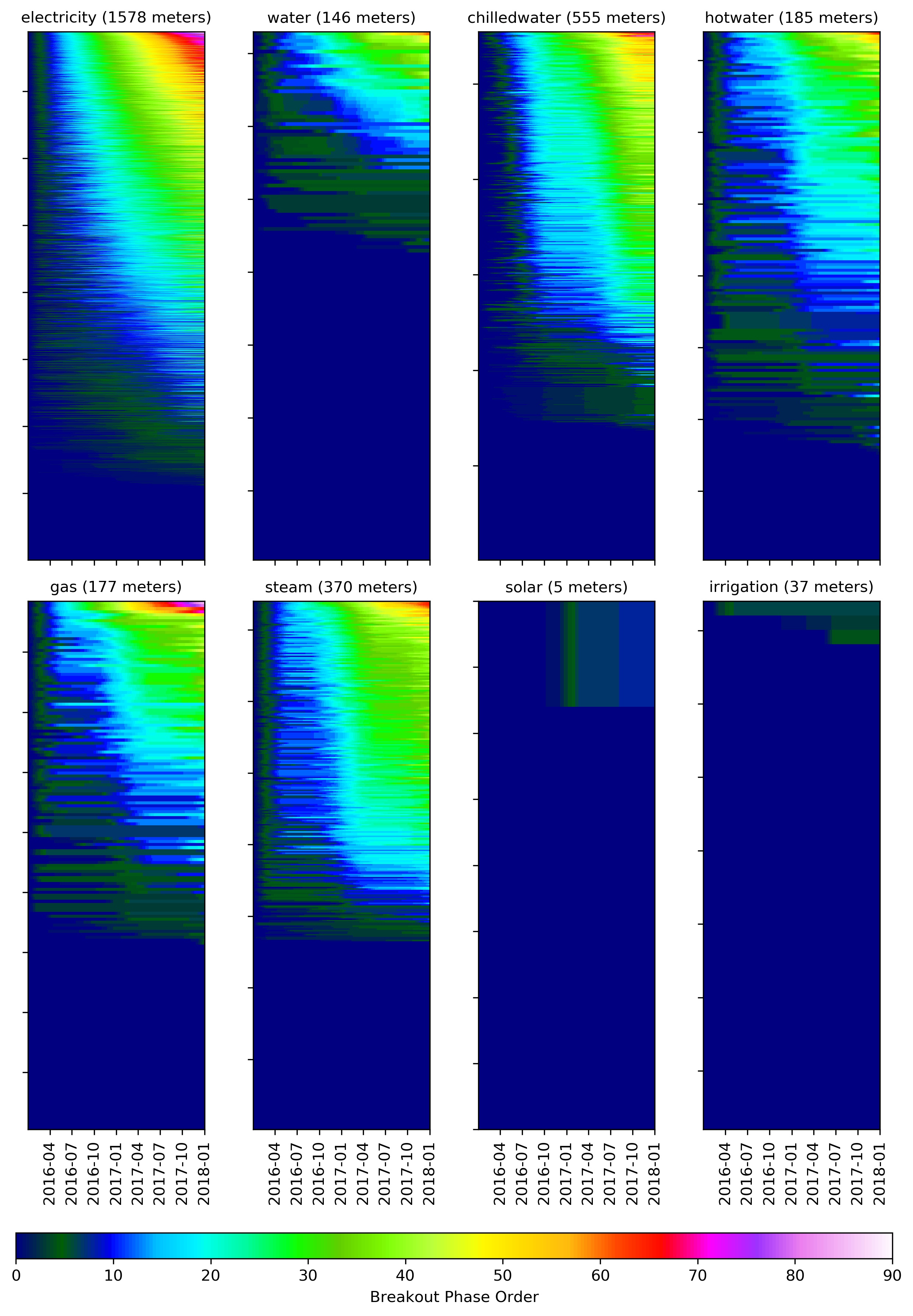}
\caption{Breakout detection heat map sorted (bottom-to-top) according to increasing number of breakouts detected. The more breakouts detected in a time-series data set, the more volatility is incurred in the data set.}
\label{tab:fig-breakout}

\end{figure*}

The final screening process shown in this publication is focused on quantifying the volatility of the time-series meter data through the use of breakout detection. A breakout is a time-series behavior that occurs when measurements have a shift from one steady-state behavior pattern to another. These shifts, or breakouts, are typically characterized by two steady states and an intermediate transition period. A breakout might be an example of a building operating in one type of schedule to another, such as commonly the case in educational buildings. For breakout detection, in this case, the Breakout Detection package developed by Twitter was used to detect the breakout shifts in an unsupervised way (\url{https://github.com/twitter/BreakoutDetection}). The critical parameter set for the model was that a steady state has to be at least 168 points long (a week) as a minimum. The resultant heat maps from this process can be seen in Figure \ref{tab:fig-breakout}. These visualizations show the volatility of consumption based on the number of breakouts detected over the time range of two years. The steam and electricity meters show a broad range of volatility, while water and gas are more consistent comparatively.

\section*{Usage Notes}
The usefulness of the BDG2 data set can be understood in the context of several applications. The most obvious is in the context of the GEPIII competition and time-series meter data prediction in general. In this section, several examples of using the data set for various applications are discussed. It should be noted that gaps or removed outliers will make an impact on the summation at the daily, weekly, or annual basis. Therefore, care should be taken when calculating metrics such as energy use intensity (EUI) at those scales without filling gaps.

\subsection*{Relationship with the GEPIII Kaggle Competition}
The first application discussed is the use of the BDG2 data set in the context of long-term data prediction. As mentioned, BDG2 includes the data that were used in the GEPIII competition on the Kaggle machine learning platform. Users of this data set can map each of the unique building ID's on the Kaggle platform, represented as an integer, with the unique ID's created in this larger data set. The documentation for that mapping can be found on a Github documentation page for the repository. Table \ref{tab:sites-tab1} includes a column that outlines which sites were used for the competition. The BDG2 includes a folder (\texttt{/data/meters/kaggle/}) that includes the data for the validation data set (2017) that matches seamlessly with the training data found on the competition website. The data contained in this folder have several differences as compared to the rest of the BDG2 data sets. The first difference is that the BDG2 data set only has timestamps in the local time zone, including the weather data. The weather data released in the Kaggle competition had a timestamp that was set to UTC, and the contestants had to come up with ways to find the right alignment for the weather data to use it properly. The other set of issues is related to several mistakes in unit conversion from the data sources and the Kaggle competition data set. Several meters that were assumed to be in kWh were in a different unit. Another issue is that several of the meters were converted from the wrong units. These mistakes have been fixed in the BDG2 data sets (\emph{raw} and \emph{cleaned}, but were left as-is in the Kaggle data set.\\

A key consideration concerning the relationship between the BDG2 and the GEPIII competition is that the third year (2018) of data from the competition is not released in this repository as some of those data are still used in the final test data (\emph{private leaderboard}) component of the competition. The competition's structure was such that the first year was released as the \emph{training} data, and the contestants were asked to produce predictions for the second and third years (2017 and 2018). In the competition, the second year was used to calculate the validation data set score (\emph{public leaderboard}), and the third year was used for the final test score (\emph{private leaderboard}). The final score test data, the third year (2018), is not released to enable users to use that year of data as the prediction objective to see how their methods match up to the contestants from the competition. Users now have two years of data from the BDG2 project to predict the third year (2018) and, therefore, it should be noted that they have an advantage over the contestants who only had access to one year of training data at the time.

\subsection*{Long-term Building Hourly Energy Prediction Model Benchmarking}
To create a curated example of meter data prediction similar to the Kaggle competition, the repository includes a well-documented instance of long-term energy prediction. These examples are described in detail in a documentation page on the repository (\url{https://github.com/buds-lab/building-data-genome-project-2/wiki/Long-term-prediction}). The example illustrated extracts various time-series feature from the meter and weather data and trains a model using one year of data to predict the following year. In this case, hourly data from 2016 is used to predict meter readings in 2017, and the accuracy as compared to ground truth is calculated using several metrics. This example is provided for users as a template for testing and incorporating their own machine learning process methods.

\subsection*{Short-term Building Hourly Energy Prediction Model Benchmarking}
The next set of examples created in the repository are similar but focus on a shorter time-scale. A large body of research exists that is focused on short-term prediction with applications more aligned with grid-scale interactions, demand response, supervisory control systems, and anomaly detection \cite{Amasyali2018-wj}. The repository provides examples of short term prediction using the data set to use one month of hourly data to predict 72 hours ahead. The detailed documentation for these examples can be found on the documentation page in the repository (\url{https://www.kaggle.com/claytonmiller/buildingdatagenomeproject2}). 

\subsection*{Building Data Genome Project 2 Kaggle Data Page}
To create an environment where users of the data set can come up with new ideas for the use of the data set, a Kaggle Data Project has been created for a community to grow ideas focused on using this data set (\url{https://www.kaggle.com/claytonmiller/buildingdatagenomeproject2}). This project is independent of the Kaggle GEPIII competition and focused on the development of kernels (or notebooks) that process the data towards various objectives. This platform enables crowd-sourcing of analysis techniques, solutions, and processes. The page has a set of \emph{Tasks} in a tab with that name that seed ideas of analysis beyond just short and long-term prediction. Some of the additional tasks outlined include time-series classification, anomaly detection, meta-data analysis, and data visualization techniques.

\section*{Code Availability}
The BDG2 data set and the custom code used for its creation and analysis is hosted in a public Github repository (\url{https://github.com/buds-lab/building-data-genome-project-2}) and its v1.0 release has been deposited in Zenodo \cite{Miller2020-fn}. This codebase includes several Jupyter notebooks with Python and R data analysis workflows that can be easily reproduced.



\section*{Acknowledgements}
The authors acknowledge the individuals who assisted in the collection of data for inclusion in this data set. This list includes (alphabetical order) Adam Boltz, Adam Keeling, Ann Lundholm, Araz Ashouri, Catherine Patton, Doug Livingston, Gerry Hamilton, Ian Lahiff, James Ball, Jonathan Roth, Justin Owen, Kian Wee Chen, Maxime St-Jacques, Nate Boyd, Saptak Dutta, and Zach Wilson. The Kaggle platform technical and advisory team, including Addison Howard and Sohier Dane, were instrumental to the launch of the GEPIII competition. The ASHRAE competition planning team of (alphabetical order) Anthony Fontanini, Chris Balbach, Jeff Haberl, and Krishnan Gowri assisted in getting the competition launched and supported by the ASHRAE organization.\\

Financial support for the development of the data set and travel support was provided by the Republic of Singapore's National Research Foundation through a grant to the Berkeley Education Alliance for Research in Singapore (BEARS) for the Singapore Berkeley Building Efficiency and Sustainability in the Tropics (SinBerBEST) program. Additional research funding was provided by the Ministry of Education (MOE) of the Republic of Singapore (R296000181133). Financial support for the  GEPIII competition monetary prizes was supported by ASHRAE. The Kaggle machine learning platform provided hosting as a non-profit competition.


\section*{Author contributions}


C.M. coordinated the creation of the data set, led the data collection and preliminary analysis for 12 of the sites, and was the lead author of the publication. A.K., P.A., and J.Y.P. each led the data collection for one site and participated in the data cleaning and pre-processing before the GEPIII competition. B.P. transformed, cleaned, and prepared the data set for publication after the GEPIII competition. Z.N., P.R., B.H., Z.S., and F.M. each contributed data for one site and provided comments to improve the data set and its use. The majority of the contribution by Z.S was made at the National Research Council Canada, Ottawa ON, Canada. All authors reviewed the manuscript and take responsibility for its content.

\section*{Competing interests}

The authors declare no competing interests.


\end{document}